\documentclass[12pt]{article}
\newcommand{\nc}{\newcommand}
\nc{\renc}{\renewcommand}

\usepackage{calc}
\usepackage{ifthen}
\usepackage{graphicx}
\usepackage{epsfig}
\usepackage{miniplot}

\nc{\half}{{\textstyle{1\over2}}}
\nc{\etal}{\mbox{\it et al. }}
\nc{\ie}{{\it i.e.}}
\nc{\eg}{{\it e.g.}}

\renc{\thefootnote}{\arabic{footnote}}
\nc{\capt}[1]{{\bf Figure.} {\small\sl #1}}

\nc{\eqs}[2]{\mbox{Eqs.~(\ref{#1},\,\ref{#2})}}
\nc{\eq}[1]{\mbox{Eq.~(\ref{#1})}}

\nc{\figs}[2]{\mbox{Figs.~(\ref{#1},\,\ref{#2})}}
\nc{\fig}[1]{\mbox{Fig~.(\ref{#1})}}

\nc{\tag}[1]{\label{#1} \marginpar{{\footnotesize #1}}}
\nc{\mtag}[1]{\label{#1} \mbox{\marginpar{{\footnotesize #1}}}}
\renc{\baselinestretch}{1.5}
\newlength{\overeqskip}
\newlength{\undereqskip}
\nc{\be}[1]{\begin{equation} \mbox{$\label{#1}$}}
\nc{\bea}[1]{\begin{eqnarray} \mbox{$\label{#1}$}}
\nc{\Section}[2]{\section{#2}\label{#1}}
\nc{\Bibitem}[1]{\bibitem{#1}}
\nc{\Label}[1]{\label{#1}}
\nc{\eea}{\vspace{\undereqskip}\end{eqnarray}}
\nc{\ee}{\vspace{\undereqskip}\end{equation}}
\nc{\bdm}{\begin{displaymath}}
\nc{\edm}{\end{displaymath}}
\nc{\dpsty}{\displaystyle}
\nc{\bc}{\begin{center}}
\nc{\ec}{\end{center}}
\nc{\ba}{\begin{array}}
\nc{\ea}{\end{array}}
\nc{\bab}{\begin{abstract}}
\nc{\eab}{\end{abstract}}
\nc{\btab}{\begin{tabular}}
\nc{\etab}{\end{tabular}}
\nc{\bit}{\begin{itemize}}
\nc{\eit}{\end{itemize}}
\nc{\ben}{\begin{enumerate}}
\nc{\een}{\end{enumerate}}
\nc{\bfig}{\begin{figure}}
\nc{\efig}{\end{figure}}
\nc{\arreq}{&\!=\!&}
\nc{\arrmi}{&\!-\!&}
\nc{\arrpl}{&\!+\!&}
\nc{\arrap}{&\!\!\!\approx\!\!\!&}
\nc{\non}{\nonumber\\*}
\nc{\align}{\!\!\!\!\!\!\!\!&&}

\def\lsim{\; \raise0.3ex\hbox{$<$\kern-0.75em
      \raise-1.1ex\hbox{$\sim$}}\; }
\def\gsim{\; \raise0.3ex\hbox{$>$\kern-0.75em
      \raise-1.1ex\hbox{$\sim$}}\; }
\nc{\DOT}{\hspace{-0.08in}{\bf .}\hspace{0.1in}}
\nc{\Laada}{\hbox {$\sqcap$ \kern -1em $\sqcup$}}
\nc\loota{{\scriptstyle\sqcap\kern-0.55em\hbox{$\scriptstyle\sqcup$}}}
\nc\Loota{{\sqcap\kern-0.65em\hbox{$\sqcup$}}}
\nc\laada{\Loota}
\nc{\qed}{\hskip 3em \hbox{\BOX} \vskip 2ex}

\nc{\real}{{\rm I \! R}}
\nc{\Z}{{\sf Z \!\!\! Z}}
\nc{\complex}{{\rm C\!\!\! {\sf I}\,\,}}
\def\bigid{\leavevmode\hbox{\small1\kern-3.8pt\normalsize1}}
\def\id{\leavevmode\hbox{\small1\kern-3.3pt\normalsize1}}
\nc{\slask}{\!\!\!/}
\nc{\bis}{{\prime\prime}}
\nc{\pa}{\partial}
\nc{\na}{\nabla}
\nc{\ra}{\rangle}
\nc{\la}{\langle}
\nc{\goto}{\rightarrow}
\nc{\swap}{\leftrightarrow}

\nc{\EE}[1]{ \mbox{$\cdot10^{#1}$} }
\nc{\abs}[1]{\left|#1\right|}
\nc{\at}[2]{\left.#1\right|_{#2}}
\nc{\norm}[1]{\|#1\|}
\nc{\abscut}[2]{\Abs{#1}_{\scriptscriptstyle#2}}
\nc{\vek}[1]{{\rm\bf #1}}
\nc{\integral}[2]{\int\limits_{#1}^{#2}}
\nc{\inv}[1]{\frac{1}{#1}}
\nc{\dd}[2]{{{\partial #1}\over{\partial #2}}}
\nc{\ddd}[2]{{{{\partial}^2 #1}\over{\partial {#2}^2}}}
\nc{\dddd}[3]{{{{\partial}^2 #1}\over
        {\partial #2 \partial #3}}}
\nc{\dder}[2]{{{d #1}\over{d #2}}}
\nc{\ddder}[2]{{{d^2 #1}\over{d {#2}^2}}}
\nc{\dddder}[3]{{d^2 #1}\over
        {d #2 d #3}}
\nc{\dx}[1]{d\,^{#1}x}
\nc{\dy}[1]{d\,^{#1}y}
\nc{\dz}[1]{d\,^{#1}z}
\nc{\dl}[1]{\frac{d\,^{#1}l}{(2\pi)^{#1}}}
\nc{\dk}[1]{\frac{d\,^{#1}k}{(2\pi)^{#1}}}
\nc{\dq}[1]{\frac{d\,^{#1}q}{(2\pi)^{#1}}}

\nc{\cc}{\mbox{$c.c.$ }}
\nc{\hc}{\mbox{$h.c.$ }}
\nc{\cf}{cf.\ }
\nc{\erfc}{{\rm erfc}}
\nc{\Tr}{{\rm Tr\,}}
\nc{\tr}{{\rm tr\,}}
\nc{\pol}{{\rm pol}}
\nc{\sign}{{\rm sign}}
\nc{\bfT}{{\bf T }}

\def\GeV{{\rm\ GeV}}
\def\MeV{{\rm\ MeV}}
\def\keV{{\rm\ keV}}
\def\TeV{{\rm\ TeV}}

\nc{\cA}{{\cal A}}
\nc{\cB}{{\cal B}}
\nc{\cD}{{\cal D}}
\nc{\cE}{{\cal E}}
\nc{\cG}{{\cal G}}
\nc{\cH}{{\cal H}}
\nc{\cL}{{\cal L}}
\nc{\cO}{{\cal O}}
\nc{\cT}{{\cal T}}
\nc{\cN}{{\cal N}}
\nc{\rvac}[1]{|{\cal O}#1\rangle}
\nc{\lvac}[1]{\langle{\cal O}#1|}
\nc{\rvacb}[1]{|{\cal O}_\beta #1\rangle}
\nc{\lvacb}[1]{\langle{\cal O}_\beta #1 |}
\nc{\bb}{\bar{\beta}}
\nc{\bt}{\tilde{\beta}}
\nc{\ctH}{\tilde{\cal H}}
\nc{\chH}{\hat{\cal H}}
\nc{\1}{\aa}
\nc{\2}{\"{a}}
\nc{\3}{\"{o}}
\nc{\4}{\AA}
\nc{\5}{\"{A}}
\nc{\6}{\"{O}}
\nc{\al}{\alpha}
\nc{\g}{\gamma}
\nc{\Del}{\Delta}
\nc{\e}{\epsilon}
\nc{\eps}{\epsilon}
\nc{\lam}{\lambda}
\nc{\om}{\omega}
\nc{\Om}{\Omega}
\nc{\ve}{\varepsilon}
\nc{\mn}{{\mu\nu}}
\nc{\vp}{\varphi}

\nc{\advp}[3]{{\it  Adv.\ in\ Phys.\ }{{\bf #1} {(#2)} {#3}}}
\nc{\annp}[3]{{\it  Ann.\ Phys.\ (N.Y.)\ }{{\bf #1} {(#2)} {#3}}}
\nc{\apl}[3]{{\it  Appl. Phys. Lett. }{{\bf #1} {(#2)} {#3}}}
\nc{\apj}[3]{{\it  Ap.\ J.\ }{{\bf #1} {(#2)} {#3}}}
\nc{\apjl}[3]{{\it  Ap.\ J.\ Lett.\ }{{\bf #1} {(#2)} {#3}}}
\nc{\app}[3]{{\it Astropart.\ Phys.\ }{{\bf #1} {(#2)} {#3}}}
\nc{\cmp}[3]{{\it  Comm.\ Math.\ Phys.\ }{{ \bf #1} {(#2)} {#3}}}
\nc{\cqg}[3]{{\it  Class.\ Quant.\ Grav.\ }{{\bf #1} {(#2)} {#3}}}
\nc{\epl}[3]{{\it  Europhys.\ Lett.\ }{{\bf #1} {(#2)} {#3}}}
\nc{\ijmp}[3]{{\it Int.\ J.\ Mod.\ Phys.\ }{{\bf #1} {(#2)} {#3}}}
\nc{\ijtp}[3]{{\it Int.\ J.\ Theor.\ Phys.\ }{{\bf #1} {(#2)} {#3}}}
\nc{\jmp}[3]{{\it  J.\ Math.\ Phys.\ }{{ \bf #1} {(#2)} {#3}}}
\nc{\jpa}[3]{{\it  J.\ Phys.\ A\ }{{\bf #1} {(#2)} {#3}}}
\nc{\jpc}[3]{{\it  J.\ Phys.\ C\ }{{\bf #1} {(#2)} {#3}}}
\nc{\jap}[3]{{\it J.\ Appl.\ Phys.\ }{{\bf #1} {(#2)} {#3}}}
\nc{\jpsj}[3]{{\it J.\ Phys.\ Soc.\ Japan\ }{{\bf #1} {(#2)} {#3}}}
\nc{\lmp}[3]{{\it Lett.\ Math.\ Phys.\ }{{\bf #1} {(#2)} {#3}}}
\nc{\mpl}[3]{{\it  Mod.\ Phys.\ Lett.\ }{{\bf #1} {(#2)} {#3}}}
\nc{\ncim}[3]{{\it  Nuov.\ Cim.\ }{{\bf #1} {(#2)} {#3}}}
\nc{\np}[3]{{\it  Nucl.\ Phys.\ }{{\bf #1} {(#2)} {#3}}}
\nc{\npps}[3]{{\it  Nucl.\ Phys.\ Proc.\ Suppl.\ }{{\bf #1} {(#2)} {#3}}}
\nc{\pr}[3]{{\it Phys.\ Rev.\ }{{\bf #1} {(#2)} {#3}}}
\nc{\pra}[3]{{\it  Phys.\ Rev.\ A\ }{{\bf #1} {(#2)} {#3}}}
\nc{\prb}[3]{{\it  Phys.\ Rev.\ B\ }{{{\bf #1} {(#2)} {#3}}}}
\nc{\prc}[3]{{\it  Phys.\ Rev.\ C\ }{{\bf #1} {(#2)} {#3}}}
\nc{\prd}[3]{{\it  Phys.\ Rev.\ D\ }{{\bf #1} {(#2)} {#3}}}
\nc{\prl}[3]{{\it Phys.\ Rev.\ Lett.\ }{{\bf #1} {(#2)} {#3}}}
\nc{\pl}[3]{{\it  Phys.\ Lett.\ }{{\bf #1} {(#2)} {#3}}}
\nc{\prep}[3]{{\it Phys.\ Rep.\ }{{\bf #1} {(#2)} {#3}}}
\nc{\prsl}[3]{{\it Proc.\ R.\ Soc.\ London\ }{{\bf #1} {(#2)} {#3}}}
\nc{\ptp}[3]{{\it  Prog.\ Theor.\ Phys.\ }{{\bf #1} {(#2)} {#3}}}
\nc{\ptps}[3]{{\it  Prog\ Theor.\ Phys.\ suppl.\ }{{\bf #1} {(#2)} {#3}}}
\nc{\physa}[3]{{\it  Physica\ A\ }{{\bf #1} {(#2)} {#3}}}
\nc{\physb}[3]{{\it  Physica\ B\ }{{\bf #1} {(#2)} {#3}}}
\nc{\phys}[3]{{\it Physica\ }{{\bf #1} {(#2)} {#3}}}
\nc{\rmp}[3]{{\it  Rev.\ Mod.\ Phys.\ }{{\bf #1} {(#2)} {#3}}}
\nc{\rpp}[3]{{\it Rep.\ Prog.\ Phys.\ }{{\bf #1} {(#2)} {#3}}}
\nc{\sjnp}[3]{{\it Sov.\ J.\ Nucl.\ Phys.\ }{{\bf #1} {(#2)} {#3}}}
\nc{\spjetp}[3]{{\it Sov.\ Phys.\ JETP\ }{{\bf #1} {(#2)} {#3}}}
\nc{\yf}[3]{{\it Yad.\ Fiz.\ }{{\bf #1} {(#2)} {#3}}}
\nc{\zetp}[3]{{\it Zh.\ Eksp.\ Teor.\ Fiz.\  }{{\bf #1}  {(#2)} {#3}}}
\nc{\zp}[3]{{\it Z.\ Phys.\ }{{\bf #1} {(#2)} {#3}}}
\nc{\ibid}[3]{{\sl ibid.\ }{{\bf #1} {#2} {#3}}}
\nc{\rf}[1]{(\ref{#1})}
\nc{\nn}{\nonumber \\*}
\nc{\bfB}{\bf{B}}
\nc{\bfv}{\bf{v}}
\nc{\bfx}{\bf{x}}
\nc{\bfy}{\bf{y}}
\nc{\vx}{\vec{x}}
\nc{\vy}{\vec{y}}
\nc{\oB}{\overline{B}}
\nc{\oI}{\overline{I}}
\nc{\oR}{\overline{R}}
\nc{\rar}{\rightarrow}
\nc{\ti}{\times}
\nc{\slsh}{\hskip-5pt/}
\nc{\sm}{Standard~Model~}
\nc{\MP}{M_{\rm Pl}}
\nc{\tp}{t_{\rm Pl}}
\nc{\ave}{\bar{E}}

\nc{\eff}{{\rm eff}}
\nc{\kk}{\vek{k}}
\nc{\pp}{{\rm p}}
\nc{\ga}{g_{a\gamma}}
\nc{\vv}{\\}
\nc{\eee}{{\bf E}}
\nc{\bbb}{{\bf B}}
\nc{\qcd}{T_{\rm QCD}}
\nc{\G}{\rm \ G}
\def\vec#1{{\bf #1}}
\def\lae{\;^{<}_{\sim} \;} \def\gae{\; ^{>}_{\sim} \;} 

\def\ell{e^{c}LL}

\begin{document}
{\title{\vskip-2truecm{\hfill {{\small \\
	\hfill \\
	}}\vskip 1truecm}
{\LARGE Unparticles: Interpretation and Cosmology
}}
%\vspace{1.2cm}
{\author{
{\sc \large John McDonald $^{1}$}\\
{\sl\small Cosmology and Astroparticle Physics Group, University of Lancaster,
Lancaster LA1 4YB, UK} }
\maketitle
%\vspace{1cm}
%\newpage
\begin{abstract}
\noindent

   We discuss the physical interpretation of unparticles and review the constraints from cosmology. Unparticles may be understood in terms of confined states of a strongly-coupled scale-invariant theory, where scale-invariance implies that the confined states have continuous masses. This picture is consistent with the observation that unparticle operators can be represented in terms of continuous mass fields. Finite results in scattering processes are obtained by compensating the infinite number of unparticle final states with an infinitesimal coupling per unparticle. As a result, unparticles are stable with respect to decay or annihilation to Standard Model particles, implying a one-way flow of energy from the Standard Model sector to the unparticle sector. The qualitative properties of unparticles, which result from their continuous mass nature, are unchanged in the case where scale-invariance is broken by a mass gap. Unparticles with a mass gap can evade constraints from astrophysical and 5th force considerations, in which case cosmology provides the strongest constraints.

%[Planning.]

\end{abstract} 
\vfil
 \footnoterule{\small  $^1$j.mcdonald@lancaster.ac.uk}   
 \newpage 
\setcounter{page}{1}                   

\section{Introduction}

    Unparticles \cite{up1,up2} represent a new possibility for the physics of a hidden sector coupled to the Standard Model (SM).
They are based on the idea of a sector which becomes strongly-coupled and scale-invariant at low energies. The form of the hidden sector is most simply a non-Abelian gauge theory with a large number of massless Dirac fermions in the fundamental representation \cite{bz}. (We will refer to this as the Banks-Zaks (BZ) sector.)  Supersymmetric QCD may also play this role \cite{fox}.
Since being proposed, there has been considerable interest in the phenomenology\footnote{We will not attempt to review here the extensive literature on unparticle phenomenology.}, astrophysics \cite{uastro,hannes,wyler,neuc} and cosmology \cite{ucosmo,dav,jm} of unparticles. Much of this work has been based on the scaling behaviour of the unparticle propagator, from which SM signatures of unparticle production can be deduced \cite{up1}. However, in order to discuss the physics of unparticles more generally, a physical picture of what unparticles are and how they are likely to behave in a given physical system is required. A particular application is cosmology, where we are concerned with an unparticle density interacting with a density of SM particles. In this case we need to understand the process of energy exchange between the two densities, which requires an understanding of how unparticles interact with each other and the SM. 

    In this paper we will review the physics of unparticles and the astrophysical and cosmological constraints on their interactions. 
The basis of our discussion is a physical interpretation in which unparticles are identified with continuous mass composite states created by a strongly-coupled BZ sector. This interpretation is consistent with the known physics of unparticles and allows their qualitative behaviour to be easily understood.

   The paper is organised as follows. In Section 2 we review the 
model of unparticles based on a strongly-coupled Banks-Zaks sector and discuss its physical interpretation
in terms of continuous mass fields. In Section 3 we review the constraints on unparticles from astrophysics and cosmology. In Section 4 we discuss the breaking of scale-invariance via the interaction of the SM Higgs with
unparticles. In Section 5 we present our conclusions.

\section{Physical Interpretation of Unparticles} 

\subsection{The Banks-Zaks Model of Unparticles} 

   In non-Abelian gauge theories coupled to massless fermions, a natural possibility is that the theory has an infra-red fixed point \cite{bz}. This occurs if the one-loop $\beta$-function is negative while the two-loop $\beta$-function is positive. An explicit example is provided by the case of an vector-like 
$SU(3)$ gauge theory coupled to $N_{F}$ massless Dirac fermions in the fundamental representation \cite{bz}.
In this case
\be{e1} \mu \frac{\partial g(\mu)}{\partial \mu} =   \beta(g)  ~\ee
where
\be{e2} \beta(g) = \left( \beta_{o} \frac{g^{3}}{16 \pi^{2}}
 + \beta_{1} \frac{g^{5}}{\left(16 \pi^{2} \right)^{2}}  \right) ~,\ee
\be{e3} \beta_{o} = -\left(11 - \frac{2}{3} N_{F}\right) ~\ee
and
\be{e4} \beta_{1} = -\left(102 –- (10 + \frac{8}{3}) N_{F}\right) ~.\ee
If  $N^{*} > N_{F} > N^{'}_{F}$, where $ N^{*} = \frac{33}{2}$ and $N_{F}^{'} = \frac{306}{38}$, then $\beta_{o}$ is negative while $\beta_{1}$ is positive. Therefore if $g$ is small at a high energy, it will increase as the renormalization scale $\mu$ decreases until the fixed point $\beta(g) = 0$ is encountered at $g = g^{*}$. This is an infra-red fixed-point of the renormalization group flow, so at $E < \Lambda_{U}$ the effective theory becomes scale-invariant. Therefore if the infra-red fixed point occurs when $g$ is non-perturbative, the BZ fields will be confined into the composite particles of a strongly-coupled scale-invariant theory at $E < \Lambda_{U}$.

\subsection{Standard Model-Unparticle Interactions}

At $E > \Lambda_{U}$, the SM and BZ particles are assumed have an 
interaction mediated by messenger fields of mass $M_{U}$ of the form 
\be{e5} {\cal L} \sim \frac{1}{M_{U}^{k}} O_{SM} O_{BZ} ~.\ee
Here $O_{SM}$ is a SM operator of mass dimension $d_{SM}$, $O_{BZ}$ is an operator of mass dimension $d_{BZ}$ made of BZ fields  and $k = d_{BZ} + d_{SM}-4$. A simple example for $O_{BZ}$ is the case of a fermion bilinear,
\be{e6} O_{BZ} = \overline{\Psi} \Psi  \;\;\; ; \;\; d_{BZ} = 3 ~.\ee
A simple example of $O_{SM}$ is
a scalar operator made of quarks \cite{up1},  
\be{e7} O_{SM} = \partial_{\mu} (\overline{u} \gamma^{\mu} (1 - \gamma_{5})t) \;\;\; ; \;\; d_{SM} = 4     ~.\ee 
Once $E < \Lambda_{U}$, the BZ sector confines into the unparticle sector. 
The BZ operators match onto unparticle operators of scaling dimension $d_{U}$ determined by the strongly-coupled theory, which create and annihilate states of the strongly-coupled scale-invariant theory, 
\be{e8} \frac{1}{M_{U}^{k}} O_{SM} O_{BZ} \rightarrow \frac{C_{U}\Lambda_{U}^{d_{BZ} -d_{U}}}{M_{U}^{k}} O_{SM} O_{U}   ~.\ee
Here $C_{U}$ is expected to be of order 1. For the SM operator of \eq{e7}, the SM-unparticle interaction in \eq{e8} becomes 
\be{e9} \frac{i \lambda}{\Lambda_{U}^{d_{U}} } \overline{u} 
\gamma_{\mu}(1- \gamma_{5}) t \partial^{\mu} O_{U}   \;\;\; ; \;\; \lambda = 
\frac{C_{U} \Lambda_{U}^{d_{BZ}}}{M_{U}^{d_{BZ}}}  ~.\ee

\subsection{Unparticles as Composite States of a Strongly-Coupled Scale-Invariant Theory}

    In order to physically understand the results of SM-unparticle interactions we need a physical interpretation of the 
unparticle states. If we consider the scale-invariant BZ theory at the infra-red fixed-point to be strongly-coupled, the BZ particles will be confined into composite states. Therefore a natural interpretation of the unparticles is that they are composite states made of confined BZ particles. Since the strongly-coupled theory producing the composite states is scale-invariant, there can be no distinct length or mass scale associated with these states. Thus composite particles with all possible masses and radii will exist. In this interpretation the unparticles will correspond to conventional particles with a continuum of mass states. This interpretation is consistent with the observation that unparticle operators can be expressed as operators creating particles with a continuous mass parameter \cite{steph,kras1,nik}.

     This continuum of mass states leads to the unconventional phenomenology and cosmology of unparticles. 
For example, the process $t \rightarrow u + {\cal U}$ was considered in \cite{up1}, where it was shown that the energy spectrum of the u-quark is continuous, in contrast with the case of production of conventional particles with a discrete mass. As reviewed below, this behaviour can be understood by expressing the unparticle operators in terms of an integral over conventional massive fields with a continuous mass parameter \cite{steph}. This is naturally interpreted as the production of composite final states with different masses, with the total inclusive decay rate to unparticles being obtained by summing over the decay to all possible final states, each of which proceeds as a conventional decay to a particle of definite mass.

\begin{figure}[h] 
                    \centering                   
                    \includegraphics[width=0.50\textwidth, angle=0]{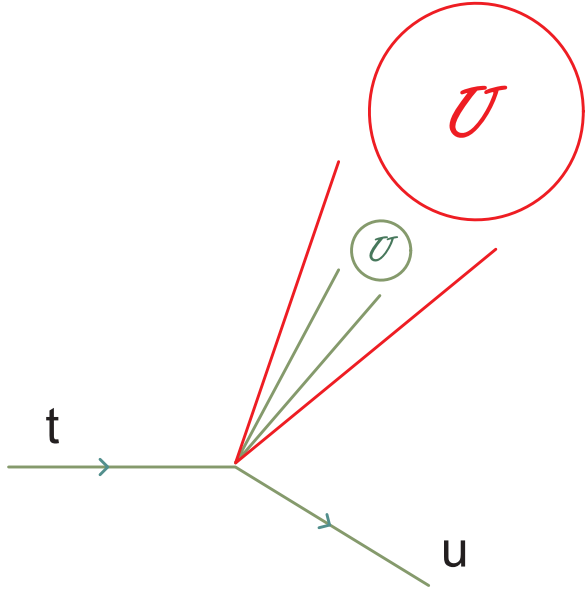}
                    \caption{\footnotesize{Schematic illustration of the decay of a t-quark to a u-quark and an unparticle. Scale-invariance implies that the unparticle can be produced with any kinematically-allowed mass and radius. The total decay rate is obtained by summing over unparticles ${\cal U}$ with different masses and radii. }}
                    \end{figure}

\section{Continuous Mass Particle Representation of Unparticle Operators}

      From the discussion of Section 2, unparticles are expected to correspond to conventional particles parameterized by a continuous mass. In \cite{steph} it was shown that scalar unparticle operators can be deconstructed as the continuum limit of scalar field operators with a mass splitting. It is natural to identify these scalar fields with continuous mass unparticles\footnote{In this case we are considering scalar unparticles. Composite states with other spins are also possible, such as spin-1 states corresponding to vector unparticles.}. 
The interpretation of unparticles as a continuum of physical particles was first proposed in \cite{kras1} and more recently emphasized in \cite{nik}. The concept of continuous mass particles was proposed much earlier in \cite{kras2}. [See also \cite{heidi,hv}.] Continuous mass particles, as a more general concept, may well open up new possibilites for phenomenology and cosmology beyond the particular example of unparticles.  
      
    We first review the equivalence of unparticle phenomenology based on scale-invariance \cite{up1} with that based on a continuum of massive particles \cite{steph,nik}. 
The correlation function of unparticle operators can be expanded in momemtum eigenstates as \cite{up1} 
\be{e10}  <0| O_{U}(x) O_{U}^{\dagger}(0) |0> = \sum_{n} <0| O_{U}(x) |p_{n}> <p_{n}|O_{U}^{\dagger}(0) |0> ~.\ee 
Translation invariance implies that 
\be{e11}  <0| O_{U}(x) |p_{n}>  = e^{-i p_{n}.x} <0| O_{U}(0) |p_{n}> ~.\ee 
Therefore         
$$ <0| O_{U}(x) O_{U}^{\dagger}(0) |0> = \sum_{n} e^{-i p_{n}.x} |<0| O_{U}(0) |p_{n}>|^{2} $$
\be{e12} \;\;\;\;\;\;  \equiv \int d^{4}P \sum_{n} \delta^{4}(P-p_{n}) e^{-i P.x} |<0| O_{U}(0) |P>|^{2} ~.\ee
Since  $\sum_{n} \delta^{4}(P-p_{n}) |<0| O_{U}(0) |P>|^{2}$ 
is a scalar function of $P^{\mu}$, it can only depend on $P^2$ and 
$\theta(P^{o})$. Therefore since the intermediate states have $p^{0} > 0$ and $p^{2} > 0$, we can write
\be{e15}   \sum_{n} \delta^{4}(P-p_{n}) |<0| O_{U}(0) |P>|^{2} = (2 \pi)^{-4} \theta(P^{o}) \theta(P^{2}) \rho(P^{2}) ~.\ee 
$\rho(P^{2})$ is the spectral function. Its form can be deduced purely from scale-invariance. Since 
\be{e16}  <0| O_{U}(x) O_{U}^{\dagger}(0) |0> = (2 \pi)^{-4} \int d^{4} P e^{-iP.x} \theta(P^{o}) \theta(P^{2}) \rho(P^{2})   ~\ee
and the LHS has scaling dimension $2 d_{U}$, it follows from scale-invariance that $\rho(P^{2})$ has scaling dimension $d_{U} - 2$. Therefore
\be{e17} \rho(P^{2})=A_{d_{U}}\left(P^{2}\right)^{ d_{U} {-} 2} ~,\ee
with $A_{d_{U}}$ a dimensionless constant which depends on the underlying strongly-coupled scale-invariant theory. 
This expression is sufficient to determine much of the phenomenology of unparticles. It allows the determination of the unparticle propagator and decay rate of SM particles to unparticles final states. Comparing \eq{e16} and \eq{e17} with the 
phase space measure for $n$ massless particles of total momentum $P$,
\be{e18} (2\pi)^{-4} A_{n} \theta(P^{0}) \theta(P^2) \left(P^{2}\right)^{n - 2 }  \;\;\; ; \;\;   A_{n} = \frac{16 \pi^{5/2}}{(2 \pi)^{2 n}} \frac{\Gamma(n + \frac{1}{2})}{\Gamma(n-1) \Gamma(2 n)} ~,\ee
it follows that, as far as the missing energy of SM particles is concerned, unparticle production looks the same as production of $d_{U}$ massless particles, where $d_{U}$ is typically non-integer \cite{up1}. The mathematical equivalence with production of a non-integer number of particles is curious but not directly significant; unparticle production differs from conventional particle production because a continuum of composite particles with different masses is produced. Each unparticle final state, once produced, will behave as a conventional particle with a well-defined mass \cite{nik}.

The scalar particle decomposition is achieved via the K\"allen-Lehmann representation 
of the unparticle propagator \cite{wein}  
\be{e19}  \int d^{4}x e^{ip.x} <0| T O_{U}(x) O_{U}^{\dagger}(0) |0> 
= \int \frac{d M^{2}}{2 \pi} \rho(M^{2}) \frac{i}{P^{2} -– M^{2} + i \epsilon}   ~.\ee 
$\rho(M^{2})$ may be expanded in terms of unparticle states $\phi_{n}$ with fixed 3-momentum 
\be{e20}  \rho(M^{2}) = 2 \pi \sum_{n} \delta(M^{2}-M_{n}^{2}) |<0| O_{U}(0) |\phi_{n}>|^{2} ~,\ee 
where scale-invariance is broken in a controlled manner by splitting the spectrum of the states as $M_{n}^{2} = \Delta^{2} n $. Substituting \eq{e20} into \eq{e19} gives
\be{e21} \int d^{4}x e^{ip.x} <0| T O_{U}(x) O_{U}^{\dagger}(0) |0> 
\rightarrow \sum_{n} \frac{i F_{n}^{2}}{P^{2} –- M_{n}^{2} + i \epsilon}  
\;\;\; ; \;\; F_{n}^{2} \equiv |<0| O_{U}(0) |\phi_{n}>|^{2} ~.\ee
Comparison with \eq{e19} after substituting \eq{e17} implies that  
\be{e23}  F_{n}^{2} = \frac{\rho(M_{n}^{2}) \Delta^{2}}{2 \pi} \equiv 
\frac{A_{d_{u}} (M_{n}^{2})^{d_{U}-2} \Delta^{2}}{2 \pi} ~.\ee
The unparticle limit is recovered as $\Delta \rightarrow 0$.   
\eq{e21} is a sum of scalar particle propagators, therefore the unparticle operator can be represented as the continuum limit of a sum of scalar field operators $\phi_{n}$ of mass $M_{n}$, 
\be{e24}  O_{U} = \sum_{n} F_{n} \phi_{n}   ~.\ee

\subsection{Decay rate to unparticles} 

        In order to make clear how SM particles interact with unparticles we consider the $t \rightarrow u + {\cal U}$ decay rate first calculated in \cite{up1} and analysed using the scalar particle decomposition in \cite{steph}.The unparticle interaction with the SM quarks, 
\be{e25} \frac{i \lambda}{\Lambda_{U}^{d_{U}} } \overline{u} 
\gamma_{\mu}(1- \gamma_{5}) t \partial^{\mu} O_{U}   + {\rm h.c.} \;\;\ ; \;\; \lambda = 
\frac{C_{U} \Lambda_{U}^{d_{BZ}}}{M_{U}^{d_{BZ}}} ~,\ee
becomes 
\be{e26}  \frac{i \lambda}{\Lambda_{U}^{d_{U}} } \overline{u} 
\gamma_{\mu}(1- \gamma_{5}) t \sum_{n} F_{n} \partial^{\mu}\phi_{n} ~.\ee
Therefore the coupling of each $\phi_{n}$ to SM particles is proportional to $\Delta$ and vanishes in the continuum limit $\Delta \rightarrow 0$. The decay rate to each $\phi_{n}$ is
\be{e27}  \Gamma(t \rightarrow u + \phi_{n}) = 
\frac{|\lambda|^{2}}{\Lambda_{U}^{2 d_{u}}} \frac{m_{t} E_{u}^{2} F_{n}^{2}}{2 \pi} ~.\ee
The total decay rate to unparticles is obtained by summing over the scalar final states in the limit $\Delta \rightarrow 0$. For each $n$ the $u$-quark energy is 
\be{e28}  E_{u} = \frac{\left(m_{t}^{2} - M_{n}^{2}\right)}{2 m_{t}}  ~.\ee
Over a small range $\Delta n$, corresponding to a range of masses $dM_{n}$  around $M_{n}$, the contribution $\Delta \Gamma$ to the total decay rate is given by
\be{e29} \Delta \Gamma = \Gamma_{n} \times \Delta n = \Gamma_{n} \times \frac{d M_{n}^{2}}{\Delta^{2}} = \frac{2 m_{t}}{\Delta^{2}} \times \Gamma_{n}  dE_{u} ~.\ee 
Therefore 
\be{e30} \frac{d \Gamma}{d E_{u}} = \frac{2 m_{t}}{\Delta^{2}}\times
\frac{|\lambda|^{2}}{\Lambda_{U}^{2 d_{u}}} \frac{m_{t} E_{u}^{2} F_{n}^{2}}{2 \pi}
\equiv \frac{2 m_{t}}{\Delta^{2}}\times
\frac{|\lambda|^{2}}{\Lambda_{U}^{2 d_{u}}} \frac{m_{t} E_{u}^{2} }{2 \pi}
\left(\frac{A_{d_{u}}}{2 \pi} \Delta^{2} (M_{n}^{2})^{(d_{u}-2)} \right)
~.\ee
The first term on the RHS is the number of scalar particle final states in the range $dE_{u}$, while the second term is the decay rate to each final state. 
The resulting energy spectrum of the u-quarks is therefore 
\be{e31} \frac{d \Gamma}{d E_{u}} = 
\frac{|\lambda|^{2}}{\Lambda_{U}^{2 d_{u}}} \frac{A_{d_{u}}m_{t}^{2} E_{u}^{2} 
(m_{t}^{2} -– 2 m_{t}E_{u})^{d_{u}-2} 
}{2 \pi^{2}}   ~.\ee 
The most important feature of \eq{e30} is that a finite total decay rate to unparticles results because the infinitesimal decay rate to each unparticle state is balanced by the infinite number of unparticle final states. 
This is quite different from conventional particle production and leads to the unusual behaviour of unparticles when coupled to the SM. Most importantly, an unparticle is unable to decay or scatter back to SM states, since the number of SM final states is finite and cannot compensate for the vanishing coupling in the $\Delta \rightarrow 0$ limit. This also implies that an unparticle energy density is stable with respect to decay to SM particles. This feature of unparticle physics is crucial for cosmology, as it implies that energy can only flow from the SM sector to the unparticle sector. As reviewed below, this one-way flow of energy into a stable unparticle energy density can impose strong cosmological constraints on unparticles \cite{jm}.

\section{Unparticles in Astrophysics and Cosmology}

   The above discussion assumes unbroken scale-invariance. In this case severe limits on $M_{U}$ are imposed by 
unparticle production in supernovae and red giants and by 5th force effects \cite{hannes,wyler,desh,neuc,dav}.  In addition, cosmology imposes limits based on the effect on Big-Bang Nucleosynthesis (BBN)  of the unparticle energy density produced by thermal SM particles \cite{dav,jm}.

\subsection{Astrophysics Limits}

In \cite{hannes} limits were imposed by considering energy loss from supernova SN1987A. Emission of unparticles via nucleon bremsstrahlung ($N + N \rightarrow N + N + {\cal U}$) was considered for a vector unparticle with interaction 
\be{as1} \frac{C_{U} \Lambda_{U}^{d_{BZ}-d_{U}}}{M_{U}^{d_{BZ}-1}} 
\overline{N} \gamma_{\mu} N O_{U}^{\mu}  ~.\ee
The rate of loss of energy from the supernova core ($ T \approx 30 \MeV$) results in an excessively shortened neutrino burst unless 
\be{as2} \frac{C_{U} \Lambda_{U}^{d_{BZ}-d_{U}} \left(30 \MeV\right)^{d_{U} - 1} }{M_{U}^{d_{BZ}-1}} \lae 3 \times 10^{-11}  ~.\ee 
With $d_{UV} = 3$, $C_{U} = 1$ and $\Lambda_{U} = 1 \TeV$ this results in 
$ M_{U} \gae 10^{8} {\rm GeV}$ ($d_{u} = 1$); $\gae 10^{7} {\rm GeV}$ ($d_{u} = 3/2$);
$  M_{U} \gae 10^{6} {\rm GeV}$ ($d_{u} = 2$). 
These bound are generally much stronger than the corresponding collider bounds, 
$M_{U} \gae 1-7.5 \TeV$ for $d_{U} = 1 - 2$ \cite{hannes}. 

   In \cite{wyler} bounds were derived from 5th force experiments and energy loss from red giants as well as from SN1987A. For a vector unparticle interaction of the form
\be{as3} \frac{C_{V} }{\Lambda_{U}^{d_{U}-1}} 
\overline{f} \gamma_{\mu} f O_{U}^{\mu}    ~,\ee 
the unparticle propagator implies that there is  
a long-range interaction mediated by unparticles which produces a potential given by
\be{as4}  V_{U} \approx \frac{\alpha_{U} B_{i} B_{j}}{r^{2 d_{U} - 1}}    ~,\ee
where $B_{i,j}$ are the baryon numbers of the objects and 
\be{as5} \alpha_{U} =  \frac{C_{V}^{2} A_{d_{u}} \Lambda_{U}^{2-2d_{U}} 
\Gamma\left(2d_{U}-2\right)}{4 \pi^{2}} ~.\ee 
\eq{as3} will also produce unparticles in the core of red-giant stars 
($T_{core} \approx 8.6 \keV$) via bremsstrahlung ($e + H^{+} \rightarrow e + H^{+} + {\cal U}$)
and the Compton process ($\gamma + e \rightarrow e + {\cal U}$). To compare with other bounds we use $C_{V} = C_{U} (\Lambda_{U}/M_{U})^{d_{BZ} -1}$. Table 1 lists the lower bounds on $M_{U}$ from 5th force experiments, energy loss from red-giants and SN1987A for the case $d_{UV} = 3$, $C_{U} = 1$ and $\Lambda_{U} = 1 \TeV$ \footnote{Based on Table 3 of \cite{wyler}.}.  
\begin{table}[h]
 \begin{center}
 \begin{tabular}{|c|c|c|c|c|}
	\hline   $M_{U}$ &  $d_{U} = 1$ &  4/3 & 5/3 & 2 \\ 
	\hline   5th Force: & $3.8 \times 10^{14} \GeV$ & $1.8 \times 10^{10} \GeV$ & $3.4 \times 10^{7} \GeV$
                          & $6.7 \times 10^{4} \GeV$\\	
      \hline   Red-Giant:& $2.2 \times 10^{10} \GeV$ & $4.2 \times 10^{8} \GeV$ & $1.4 \times 10^{7} \GeV$
                          & $5.1 \times 10^{5} \GeV$\\
      \hline   SN1987A:   & $3.2 \times 10^{7} \GeV$ & $3.6 \times 10^{6} \GeV$ & $4.5 \times 10^{5} \GeV$
                          & $5.5 \times 10^{4} \GeV$\\
      \hline     
 \end{tabular}
 \caption{\footnotesize{5th Force, Red-Giant Cooling and SN1987A Bounds on Unparticles \cite{wyler}.}}  
 \end{center}
 \end{table}
\noindent From this we see that 5th force experiments impose particularly strong lower bounds on $M_{U}$ in the case where unparticles have unbroken scale invariance. In this case it is unlikely that unparticles will be detected at colliders. 

    These bounds can be evaded if scale-invariance of the unparticles is sufficiently broken. In the case where they have a mass greater than $30 \MeV$, unparticles are too massive to be produced in supernovae or red-giants and will induce only short-range forces. In this case the most important constraints on $M_{U}$ will come from cosmology.

\subsection{Cosmological Constraints}

       Cosmological constraints on unparticles were first considered in \cite{dav}. This analysis was based on a number of assumptions, in particular that unparticles can come into thermal equilibrium with SM particles and that decoupling of an unparticle density before the QCD phase transition can dilute it sufficiently to evade BBN constraints. However, as reviewed below, establishment of a thermal equilibrium between SM particles and unparticles, when interpreted as continuous mass particles, is non-trivial due to their unconventional production process. Moreover, unless the unparticle density is less than the SM particle density prior to the QCD transition, dilution will not be sufficient to evade BBN constraints \cite{jm}.

          In the unparticle limit, $\Delta \rightarrow 0$, the coupling of each unparticle to SM particles vanishes. 
In any process where the unparticles decay or annihilate to SM states, the number of SM final states is finite. Thus as $\Delta \rightarrow 0$ there is no compensating increase in the number of final states, unlike the case of SM particles scattering to unparticles. Therefore the rate of production of SM particles from unparticles is zero. As a result it is not possible for unparticles to come into thermal equilibrium with SM particles. 

     If the unparticle production rate from SM particles, $\Gamma$, is larger than the expansion rate, $H$, energy will flow in one direction from the SM density to the unparticle density. The temperature $T$ of the SM particles will then drop until the condition $\Gamma < H$ is satisfied. Thus if $\Gamma > H$ is satisfied at any time in the early Universe, a large and stable unparticle density will be created. As discussed below, this will generally violate BBN constraints. Therefore a fundamental condition for a successful unparticle cosmology is that $\Gamma < H$ at all temperatures for which the unparticle description is correct, $3T \lae \Lambda_{U}$. 

   For the simple interaction between scalar unparticles and quarks described by \eq{e9}, the decay rate of thermal quarks $q^{'} \rightarrow q + {\cal U}$ with energy $E \approx 3T$ is estimated to be \cite{jm}  
\be{as6} \Gamma \approx \frac{ C_{U}^{2} A_{d_{U}} }{24 \pi^{2}}
 \frac{ \Lambda_{U}^{2(d_{BZ} - d_{U})} }{d_{U}(d_{U}^{2} - 1)M_{u}^{2 d_{BZ}}}  
\frac{m_{q^{'}}^{2 (d_{U} + 1)} }{T}    \;\;\;\;  ;   \;\;  T < T_{EW}
~\ee 
\be{as7} \Gamma \approx \frac{ C_{U}^{2} A_{d_{U}} }{24 \pi^{2}}
 \frac{ \Lambda_{U}^{2(d_{BZ} - d_{U})} T^{2 d_{U} + 1}    }{d_{U}(d_{U}^{2} - 1)M_{u}^{2 d_{BZ}}}  
%\lambda_{q^{'}}^{2 (d_{U} + 1) } 
\;\;\;\;  ;   \;\;  T > T_{EW}
~.\ee 
For $T < T_{EW}$, the constraint $\Gamma < H$ is strongest at the lowest value of $T$ for which there are relativistic t-quarks, $T \approx m_{t}/3$. 
From \eq{as6} we then obtain an lower bound on $M_{U}$,     
\be{as8}  M_{U}^{2 d_{BZ}}  > \frac{9}{8 \pi^2} \frac{k_{1}}{k_{T}}   
\frac{m_{t}^{2 d_{U}-1}M\Lambda_{U}^{2(d_{BZ} - d_{U})}}{d_{U}(d_{U}^{2} - 1)}  \;\;\; ; \;\; k_{1} =  C_{U}^{2} A_{d_{U}} ~,\ee
where during radiation-domination $H = k_{T}T^2/M$ with $k_{T} = (\pi^2 g(T)/90)^{1/2}$.
The lower bounds on $M_{U}$ as a function of $\Lambda_{U}$ for the case $d_{BZ} = 3$ are shown in Figure 2. From \eq{as8} we conclude that for $1.1 \leq d_{U} \leq 2$ and $2 \leq d_{BZ} \leq 4$, the lower bound on $M_{U}$ is in the range 20-2600 TeV \cite{jm}.  

For $ T > T_{EW}$, the condition $\Gamma < H$ combined with \eq{as7} gives an upper bound on the temperature of the 
radiation-dominated Universe
\be{as9}   T^{2 d_{U} - 1} < 24 \pi^{2}
\frac{k_{T}}{k_{1}} 
\frac{ d_{U}(d_{U}^{2} - 1) M_{u}^{2 d_{BZ}}}{ \Lambda_{U}^{2(d_{BZ} - d_{U})}M }  ~.\ee
The upper bound on the temperature for the case $\Lambda_{U} = 1 \TeV$ and $d_{BZ} = 3$ is shown in Figure 3. (The upper bound is stronger for larger $\Lambda_{U}$.) For most of the parameter space the upper bound is greater than the largest temperature for which the
unparticle description is valid, $E \approx 3T \lae \Lambda_{U}$. For larger temperatures, interactions of SM fields will produce BZ particles rather than unparticles. In this case a thermal equilibrium can be established since the BZ particles are conventional particles. The condition for equilibrium can be estimated from the decay and inverse decay rate of SM quarks to 
BZ particles. (A similiar rate is obtained for scattering  $q^{'} q \leftrightarrow \overline{\Psi} \Psi$.)
 With $O_{BZ}$ corresponding to a bilinear of BZ fermions $\overline{\Psi}\Psi$, the interaction with the SM quarks becomes 
\be{as10}        \frac{1}{M_{U}^{k}}O_{SM}O_{BZ}  \rightarrow \frac{ \partial_{\mu}(\overline{q} \gamma_{\mu}\left(1- \gamma_{5}\right) q^{'} ) \overline{\Psi}\Psi }{M_{U}^{3}}              ~.\ee 
By dimensional considerations the decay rate for $q^{'} \rightarrow q \overline{\Psi} \Psi$ is,  
\be{as11} \Gamma \approx \frac{1}{\left(8 \pi\right)^3} \frac{T^{7}}{M_{U}^{6}}            ~.\ee
The condition $\Gamma < H$ then implies that 
\be{as12} T \; \lae \; 1.8 \; \left(\frac{M_{U}}{100 \TeV}\right)^{6/5} \TeV    ~.\ee 
If this bound is not satisfied,  
thermal equilibrium will be established with the BZ sector, 
resulting in a large BZ sector energy density. 
For the case of the simplest BZ theory, with an $SU(3)$ gauge group and $17 > N_{F} >8$ Dirac fermions in the 
fundamental representation of $SU(3)$ \cite{bz}, the number of thermal degrees of freedom is at least $g(T) = 8 \times 2 + 7/8 \times 4 \times 3 \times 8 = 100.0$, which is similar to the number of degrees of freedom in the SM, $g(T) = 106.75$. Therefore in thermal equilibrium, $\rho_{BZ} \approx \rho_{SM}$. Once $T$ drops below the upper limit in \eq{as12}, the SM and BZ sectors decouple. As the BZ fields lose energy via expansion they will evolve into unparticles, leaving an energy density $\rho_{U} \approx \rho_{SM}$ in the unparticle sector. 

    It has been suggested that the QCD transition can dilute the decoupled unparticle/BZ density \cite{dav}. However, this is ineffective unless the density in unparticles prior to the QCD transition satisfies $\rho_{U} < 0.15 \rho_{SM}$. BBN allows at most an additional 6$\%$
contribution to the energy density at nucleosynthesis \cite{bbn}. By scale-invariance the unparticle density is expected to evolve with expansion according to $\rho_{U} \propto a^{-4}$. Therefore the 
unparticle energy density after the QCD transition is 
\be{as12b} \rho_{U}(T_{BBN}) = \left(\frac{a}{a_{BBN}}\right)^{4} \rho_{U}(T) = \left(\frac{g_{*}(T_{BBN})}{g_{*}(T)}\right)^{4/3} \left(\frac{T_{BBN}}{T}\right)^{4} \rho_{U}(T) ~,\ee
where $g_{*}(T)$ is the effective number of degrees of freedom for the conserved entropy.  
The SM radiation density, $\rho_{SM}$, is proportional to $g(T)T^{4}$. Therefore 
\be{as12a}  \left(\frac{\rho_{U}}{\rho_{SM}}\right)_{BBN} = 
\left(\frac{g_{*}(T_{BBN})}{g_{*}(T)}\right)^{4/3} \left(\frac{g(T)}{g(T_{BBN})}\right)
\left(\frac{\rho_{U}}{\rho_{SM}}\right)_{T}     ~.\ee
With $g(T_{BBN}) = 3.36$ and $g_{*}(T_{BBN}) = 3.90$, corresponding to photons plus out-of-equilibrium neutrinos, and $g(T) = g_{*}(T) = 106.75$, corresponding to all SM particles, the maximum suppression of $\rho_{U}/\rho_{SM}$ is by a factor 0.39. Therefore in order to have $\left(\frac{\rho_{U}}{\rho_{SM}}\right)_{BBN} < 0.06$ we require that $\left(\frac{\rho_{U}}{\rho_{SM}}\right)_{T} < 0.15$ at $T > T_{EW} \gg T_{QCD}$. Thus dilution of the unparticle density after decoupling of the SM is generally ineffective. The unparticle density cannot be sufficiently diluted even in the case where decoupling occurs in the BZ regime, $T \gae  \Lambda_{U}/3$. Therefore an upper limit on the temperature of the Universe is a general feature of the interaction of the SM with an unparticle/BZ sector.

\begin{figure}[h] 
                    \centering                   
                    \includegraphics[width=0.50\textwidth, angle=-90]{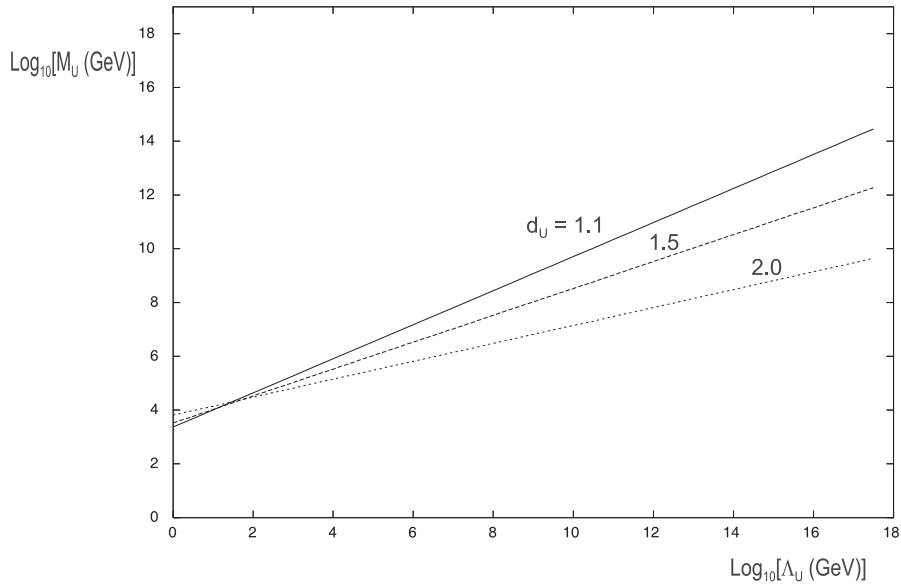}
                    \caption{\footnotesize{Lower bound on the messenger mass $M_{U}$ as a function of $\Lambda_{U}$ for $d_{U} = $ 1.1, 1.5 and 2.0. }}
                    \end{figure}

\begin{figure}[h] 
                    \centering                   
                    \includegraphics[width=0.50\textwidth, angle=-90]{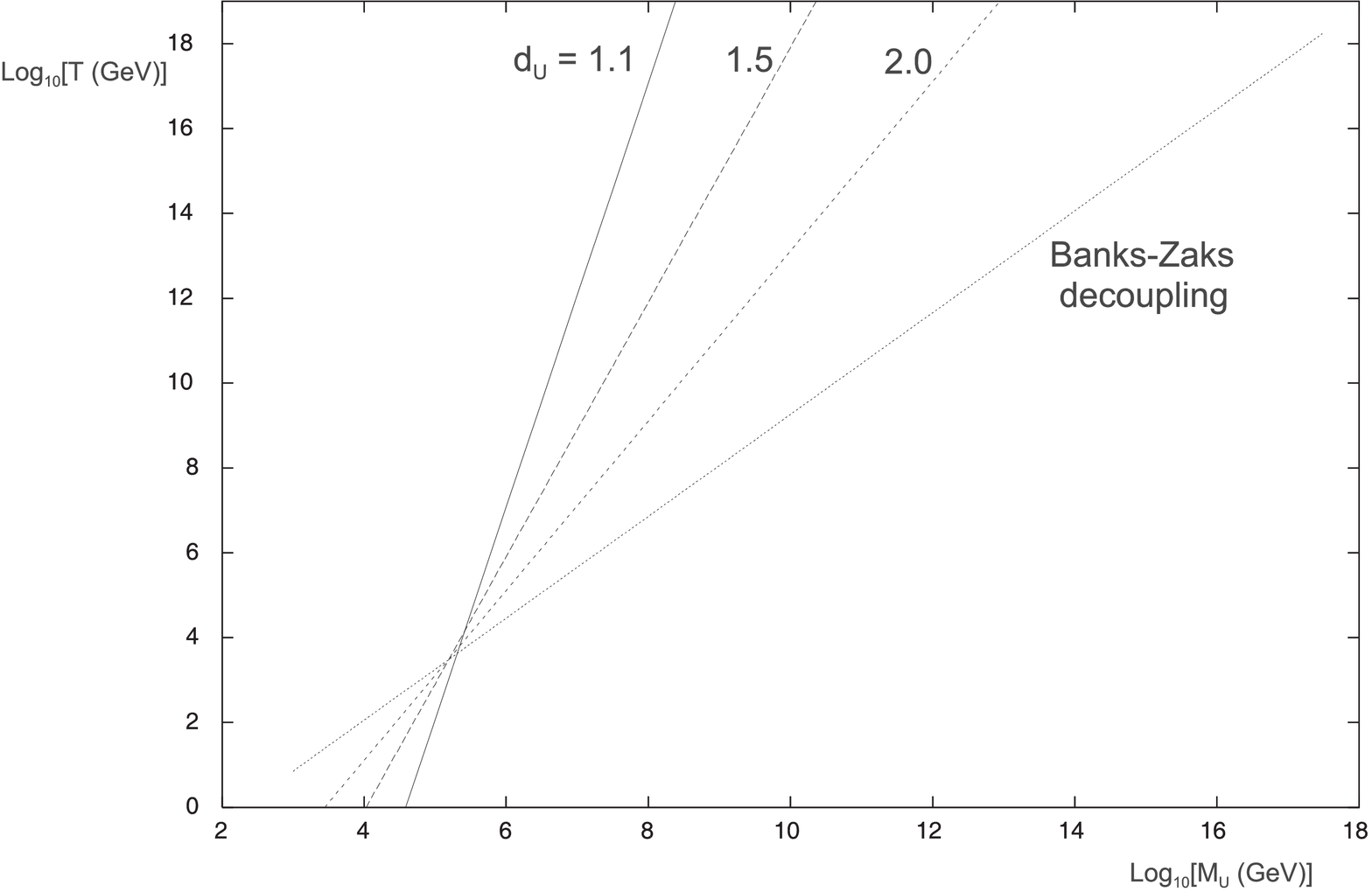}
                    \caption{\footnotesize{Upper bound on the temperature of the Standard Model for $\Lambda_{U} = 1 \TeV$ and $d_{u} = $ 1,1, 1.5 and 2.0. 
The upper bound from Banks-Zaks decoupling is also shown. }}
                    \end{figure}

 These cosmological limits are dependent on the stability of the 
BZ/unparticle sector with respect to decay to SM particles. The origin of stability is different in the BZ and unparticle limits. Stability of the BZ particles follows from their charge with respect to the BZ gauge field. Since the SM fields are not charged under this gauge group, there is no possibility of the decay of a BZ particle to SM particles and only annihilation of BZ particles is possible. Once the BZ fields are confined into unparticles, stability follows from the continuous mass nature of the unparticles, which implies a vanishing coupling per unparticle if the production rate of unparticles by SM particles is finite.

\section{Breaking scale-invariance}

         Unparticles are consistent with astrophysics, 5th force constraints and cosmology if $M_{U}$ is large enough.  To evade supernova constraints with lower values of $M_{U}$ the minimum mass of the unparticles must be greater than 30MeV. The physics of unparticles will then crucially depend on how the scale-invariance is broken. There are broadly two possibilities: (i) via a discrete mass per unparticle or (ii) via a mass gap. 

    In the case where each unparticle has a discrete mass, for example where $\Delta$ is finite, there will be a finite decay and annihilation rate of unparticles to SM particles. In this case the unparticle density will be unstable. 

   However, if scale-invariance is broken only via interactions with the SM, then it is likely that a mass gap 
will be generated, corresponding to a minimum mass in the continuum of mass states \cite{del,del2} \footnote{An essentially equivalent approach, in which the spectral function has modes with energy less than $\mu$ removed, was considered in \cite{fox,barger}}.  
The simplest example is where there is an interaction between the Higgs bi-linear and BZ fields \cite{del},
\be{as13} \frac{1}{M_{U}^{d_{BZ}-2}} O_{BZ} H^{\dagger} H   ~.\ee 
At $E < \Lambda_{U}$ this produces an interaction with the scalar unparticle operator of the form 
\be{as14}  C_{U} \left(\frac{\Lambda_{U}}{M_{U}}\right)^{d_{BZ} -2}
 \Lambda_{U}^{2 -– d_{u}} |H|^{2} O_{U}  = \kappa_{U} |H|^2 O_{U} ~.\ee
In terms of the expansion of unparticles in scalar fields, this becomes
\be{as15} \kappa_{U} |H|^2 \sum_{n} F_{n} \phi_{n}   + \frac{1}{2} \sum_{n} M_{n}^{2} \phi_{n}^2  ~.\ee
Once the Higgs develops a vacuum expectation value $<|H|^2> = v^2/2$, a term linear in each $\phi_{n}$ 
appears in the potential. This results in an expectation value for each $\phi_{n}$
\be{as16}  < \phi_{n}>  = - \frac{\kappa_{U} v^2 F_{n}}{2 M_{n}^2}      ~.\ee
This leads to a problem in the unparticle limit. Since $F_{n} \propto \Delta$ while $M_{n}^2 \propto \Delta^2$, it follows that $< \phi_{n}>$ diverges as $\Delta \rightarrow 0$. To evade this problem, an additional interaction must be added 
to the model. By generating a mass gap 
\be{as17} M_{n}^{2} \rightarrow M_{n}^{2} + m_{gap}^2   ~,\ee
the divergence in the unparticle limit can be avoided. This can be achieved by adding an interaction of the form \cite{del}
\be{as18}  \zeta |H|^2 \sum_{n} \phi_{n}^{2}        ~.\ee
This generates a mass gap with $m_{gap}^2 = \zeta v^2$. \eq{as15} and \eq{as18} also induce a mixing between the unparticles and Higgs scalar. However, the mixing angle is generally proportional to $F_{n}$ or $<\phi_{n}>
$ and therefore vanishes in the limit $\Delta \rightarrow 0$ for each individual unparticle 
state $\phi_{n}$. 

        The mass gap and unparticle-Higgs mixing will 
generally have an effect on the Higgs production rate and decay width \cite{del}. For cosmology the most important result is that when scale-invariance is broken via a mass gap the unparticles will still be completely stable. This follows because the unparticles still have a continuous mass parameter and so each unparticle has an infintesimal probability of being produced by SM particles in the limit $\Delta \rightarrow 0$. (The unparticle couplings to the SM quarks are unaffected by the coupling to the Higgs.) Once an unparticle is produced, it will have vanishing coupling to SM states and vanishing mixing with the Higgs. Therefore each $\phi_{n}$ is stable in the unparticle limit. The cosmology of unparticles with a mass gap will therefore be unaltered at energies larger than $m_{gap}$. One-way energy flow to the unparticle sector will occur as before if $\Gamma > H$, resulting in the same cosmological bounds.

\section{Conclusions}

  Unparticles represent a new form of hidden sector coupled to the Standard Model. We have summarized a physical interpretation of unparticles that allows their behaviour to be more directly understood. In this interpretation unparticles are identified with the composite particles of an underlying strongly-coupled scale-invariant theory. A consequence of scale-invariance is that the composite particles will have a continuous mass parameter. The physics of unparticles is therefore equivalent to the physics of continuous mass particles. 

   An unparticle produced by interaction with Standard Model particles will behave as a conventional particle with a discrete mass. This reflects a curious asymmetry in the physics of unparticles, depending on whether we are considering the inclusive production of unparticles or focusing on a particular unparticle final state. The former requires consideration of an infinite number of final states each with an infinitesimal coupling to the Standard Model particles, while the latter looks like a conventional particle with a well-defined mass and zero coupling to the Standard Model. 

     The unusual phenomenology of unparticles is a consequence of the infinite number of possible unparticle states, which compensates the vanishing coupling of each unparticle to Standard Model particles in the continuous mass limit. As a result, an unparticle will be completely stable with respect to decay to Standard Model particles once produced. Therefore energy can only flow in one direction from the Standard Model sector to the unparticle sector, with the resulting unparticle density being stable. At higher energies this stability is extended into the Banks-Zaks sector, where the gauge charge of the Banks-Zaks particles prevents their decay into Standard Model particles. 

      Astrophysical and 5th-force bounds impose strong constraints on the interaction between unparticles and Standard Model particles, which essentially exclude any possibility of observing unparticles at colliders. However, if scale-invariance is broken, such that the unparticles have a 
mass greater than 30 MeV, then these constraints may be evaded. The resulting phenomenology and cosmology of unparticles will then depend crucially on how scale-invariance is broken. If each unparticle acquires a discrete mass then unparticles will behave as conventional particles, with a finite decay rate to Standard Model particles and an unstable energy density. On the other hand, if unparticles have a mass gap, corresponding to a continuum of masses with a minimum mass, then the cosmology of unparticles will be qualitatively unchanged: unparticles will still be stable with respect to decay to the Standard Model in order to have a finite unparticle production rate, while energy will flow one-way from the Standard Model energy density to a stable unparticle density. It may be significant that mass gap breaking of scale-invariance appears to occur naturally as a result of the interaction of unparticles with the Standard Model Higgs.     
   
    Cosmology provides the strongest bounds on unparticle interactions in the case where scale-invariance and the unparticle density is stable at the time of Big-Bang nucleosynthesis. The requirement that the scalar unparticle energy density is consistent with Big-Bang nucleosynthesis constrains the messenger mass $M_{U}$ to be greater than 20-2400 TeV when $1.1 \leq d_{U} \leq 2$, $2 \leq d_{BZ} \leq 4$ and $\Lambda_{U} \gae  1 \TeV$. Moreover, the temperature of the Standard Model sector has an upper bound, which can be as low as 1-10 TeV for $M_{U}$ near its lower bound. 
These cosmological bounds on suggest that it will be difficult for collider experiments to detect evidence of unparticles in the absence of a mechanism for destabilizing the unparticle density before nucleosynthesis.

\section*{Acknowledgements} 

This work was supported by the European Union through the Marie Curie Research and Training Network "UniverseNet" (MRTN-CT-2006-035863) and by STFC (PPARC) Grant PP/D000394/1.

%\newpage


\begin{thebibliography}{50}

\bibitem{up1} 
  H.~Georgi,
  %``Unparticle Physics,''
  Phys.\ Rev.\ Lett.\  {\bf 98}, 221601 (2007)
  [arXiv:hep-ph/0703260].
  %%CITATION = PRLTA,98,221601;%%


\bibitem{up2}
  H.~Georgi,
  %``Another Odd Thing About Unparticle Physics,''
  Phys.\ Lett.\  B {\bf 650}, 275 (2007)
  [arXiv:0704.2457 [hep-ph]].
  %%CITATION = PHLTA,B650,275;%%

\bibitem{bz}
  T.~Banks and A.~Zaks,
  %``On The Phase Structure Of Vector-Like Gauge Theories With Massless
  %Fermions,''
  Nucl.\ Phys.\  B {\bf 196}, 189 (1982).
  %%CITATION = NUPHA,B196,189;%%


\bibitem{fox}
  P.~J.~Fox, A.~Rajaraman and Y.~Shirman,
  %``Bounds on Unparticles from the Higgs Sector,''
  Phys.\ Rev.\  D {\bf 76}, 075004 (2007)
  [arXiv:0705.3092 [hep-ph]].
  %%CITATION = PHRVA,D76,075004;%%


\bibitem{uastro}
  
  Y.~Liao and J.~Y.~Liu,
  %``Long-ranged spin-spin interaction of electron from unparticle    
  arXiv:0706.1284 [hep-ph];
  %%CITATION = ARXIV:0706.1284;%%
  N.~G.~Deshpande, S.~D.~H.~Hsu and J.~Jiang,
  %``Long range forces and limits on unparticle interactions,''
  arXiv:0708.2735 [hep-ph];
  %%CITATION = ARXIV:0708.2735;%%
  P.~K.~Das,
  %``Unparticle effects in Supernovae cooling,''
  arXiv:0708.2812 [hep-ph].
  %%CITATION = ARXIV:0708.2812;%%

\bibitem{hannes}
S.~Hannestad, G.~Raffelt and Y.~Y.~Y.~Wong,
  %``Unparticle constraints from SN1987A,''
  arXiv:0708.1404 [hep-ph].
  %%CITATION = ARXIV:0708.1404;%%

\bibitem{wyler} 
A.~Freitas and D.~Wyler,
  %``Astro Unparticle Physics,''
  arXiv:0708.4339 [hep-ph].
  %%CITATION = ARXIV:0708.4339.%%

\bibitem{desh}
  N.~G.~Deshpande, S.~D.~H.~Hsu and J.~Jiang,
  %``Long range forces and limits on unparticle interactions,''
  Phys.\ Lett.\  B {\bf 659} (2008) 888
  [arXiv:0708.2735 [hep-ph]].
  %%CITATION = PHLTA,B659,888;%%

\bibitem{neuc} 
M.~C.~Gonzalez-Garcia, P.~C.~de Holanda and R.~Zukanovich Funchal,
  %``Constraints from Solar and Reactor Neutrinos on Unparticle Long-Range
  %Forces,''
  arXiv:0803.1180 [hep-ph].
  %%CITATION = ARXIV:0803.1180;%%


\bibitem{ucosmo}

  I.~Lewis,
  %``Cosmological and Astrophysical Constraints on Tensor Unparticles,''
  arXiv:0710.4147 [hep-ph];
  %%CITATION = ARXIV:0710.4147;%%

  G.~L.~Alberghi, A.~Y.~Kamenshchik, A.~Tronconi, G.~P.~Vacca and G.~Venturi,
  %``Cosmological Unparticle Correlators,''
  Phys.\ Lett.\  B {\bf 662}, 66 (2008)
  [arXiv:0710.4275 [hep-th]];
  %%CITATION = PHLTA,B662,66;%%

  S.~L.~Chen, X.~G.~He, X.~P.~Hu and Y.~Liao,
  %``Thermal Unparticles: A New Form of Energy Density in the Universe,''
  arXiv:0710.5129 [hep-ph];
  %%CITATION = ARXIV:0710.5129.%%

  T.~Kikuchi and N.~Okada,
  %``Unparticle Dark Matter,''
  arXiv:0711.1506 [hep-ph];
  %%CITATION = ARXIV:0711.1506;%%

  H.~Collins and R.~Holman,
  %``Unparticles and inflation,''
  arXiv:0802.4416 [hep-ph];
  %%CITATION = ARXIV:0802.4416;%%

  Y.~Gong and X.~Chen,
  %``Cosmological constraint on unparticle dark matter,''
  arXiv:0803.3223 [astro-ph].
  %%CITATION = ARXIV:0803.3223;%%

\bibitem{dav}
  H.~Davoudiasl,
  %``Constraining Unparticle Physics with Cosmology and Astrophysics,''
  arXiv:0705.3636 [hep-ph].
  %%CITATION = ARXIV:0705.3636;%%

\bibitem{jm} 
  J.~McDonald,
  %``Cosmological Constraints on Unparticles,''
  arXiv:0709.2350 [hep-ph].
  %%CITATION = ARXIV:0709.2350;%%


\bibitem{steph}
  M.~A.~Stephanov,
  %``Deconstruction of Unparticles,''
  Phys.\ Rev.\  D {\bf 76}, 035008 (2007)
  [arXiv:0705.3049 [hep-ph]].
  %%CITATION = PHRVA,D76,035008;%%

\bibitem{kras1} N.~V.~Krasnikov,
  %``Unparticle as a field with continuously distributed mass,''
  arXiv:0707.1419 [hep-ph].
  %%CITATION = ARXIV:0707.1419;%%

\bibitem{nik}
  H.~Nikolic,
  %``Unparticle as a particle with arbitrary mass,''
  arXiv:0801.4471 [hep-ph].
  %%CITATION = ARXIV:0801.4471;%%


\bibitem{kras2} 
  N.~V.~Krasnikov,
  %``Higgs boson with continuously distributed mass,''
  Phys.\ Lett.\  B {\bf 325} (1994) 430.
  %%CITATION = PHLTA,B325,430;%%


\bibitem{heidi}  
  J.~J.~van der Bij and S.~Dilcher,
  %``HEIDI and the unparticle,''
  arXiv:0707.1817 [hep-ph];
  %%CITATION = ARXIV:0707.1817;%%

A.~Ferroglia, A.~Lorca and J.~J.~van der Bij,
  %``The Z' reconsidered,''
  Annalen Phys.\  {\bf 16} (2007) 563
  [arXiv:hep-ph/0611174];
  %%CITATION = ANPYA,16,563;%%


  J.~J.~van der Bij,
  %``The minimal non-minimal standard model,''
  Phys.\ Lett.\  B {\bf 636}, 56 (2006)
  [arXiv:hep-ph/0603082].
  %%CITATION = PHLTA,B636,56;%%

  

\bibitem{hv} 
  M.~J.~Strassler,
  %``Why Unparticle Models with Mass Gaps are Examples of Hidden Valleys,''
  arXiv:0801.0629 [hep-ph].
  %%CITATION = ARXIV:0801.0629;%%


\bibitem{wein} S.Weinberg, {\it The Quantum Theory of Fields I}, C.U.P (1995). 

\bibitem{bbn}
 J.~P.~Kneller and G.~Steigman,
  %``BBN and CMB constraints on dark energy,''
  Phys.\ Rev.\  D {\bf 67}, 063501 (2003)
  [arXiv:astro-ph/0210500].
  %%CITATION = PHRVA,D67,063501;%%


\bibitem{del} 
 A.~Delgado, J.~R.~Espinosa and M.~Quiros,
  %``Unparticles-Higgs Interplay,''
  arXiv:0707.4309 [hep-ph].
  %%CITATION = ARXIV:0707.4309;%%

\bibitem{del2} 
  A.~Delgado, J.~R.~Espinosa, J.~M.~No and M.~Quiros,
  %``The Higgs as a Portal to Plasmon-like Unparticle Excitations,''
  JHEP {\bf 0804}, 028 (2008)
  [arXiv:0802.2680 [hep-ph]];
  %%CITATION = JHEPA,0804,028;%%
  A.~Delgado, J.~R.~Espinosa, J.~M.~No and M.~Quiros,
  %``Phantom Higgs from Unparticles,''
  arXiv:0804.4574 [hep-ph].
  %%CITATION = ARXIV:0804.4574;%%


\bibitem{barger}
  V.~Barger, Y.~Gao, W.~Y.~Keung, D.~Marfatia and V.~N.~Senoguz,
  %``Unparticle physics with broken scale invariance,''
  Phys.\ Lett.\  B {\bf 661}, 276 (2008)
  [arXiv:0801.3771 [hep-ph]].
  %%CITATION = PHLTA,B661,276;%%


\end{thebibliography}
\end{document}